\documentclass[a4paper,oneside,final,notitlepage,onecolumn,12pt]{article}
\usepackage{amsfonts}
\usepackage{epsf}
\usepackage{graphicx}% Include figure files
\usepackage{amssymb,eso-pic}
\usepackage{latexsym}
\usepackage{tabularx}
\usepackage{amsxtra,accents}
\usepackage{hyperref}
\usepackage{t1enc}
\usepackage{amsmath}
\usepackage[T1]{fontenc}
\usepackage{amsmath}
\usepackage{authblk}

\usepackage{framed}
\usepackage{enumerate}
\usepackage{bm} 
\usepackage{mathrsfs}
\usepackage{amsthm}
\usepackage{latexsym}
\usepackage{tikz}
\usepackage{mathtools}
\DeclareGraphicsExtensions{.jpg,.jpeg,.pdf,.png,.eps}

\usepackage{ifpdf,epsfig,array,amsmath,amssymb}

\usepackage{psfrag}
\psfrag{na}{\footnotesize $n^a$}   
\psfrag{ta}{\footnotesize $\tau^a$}   
\psfrag{vna}{\footnotesize $\check{n}^a$}
\psfrag{vNa}{\footnotesize $\check{N}^a$}
\psfrag{scrW}{\footnotesize $\mathscr{W}_{\rho_0}$}
\psfrag{t=t1}{\footnotesize $\Sigma_{\tau_1}$}
\psfrag{t=t2}{\footnotesize $\Sigma_{\tau_2}$}
\psfrag{r=all}{\footnotesize $\rho=const$}
\psfrag{hna}{\footnotesize $\widehat{n}^a$}
\psfrag{tna}{\footnotesize $\tilde{n}^a$}
\psfrag{hab,Kab}{\footnotesize $h_{ab}, K_{ab}$}
\psfrag{thab,tKab}{\footnotesize $\tilde h_{ab}, \tilde K_{ab}$}
\psfrag{hhab,hKab}{\footnotesize $\widehat h_{ab}, \widehat K_{ab}$}
\psfrag{chab,cKab}{\footnotesize $\check{h}_{ab}, \check{K}_{ab}$}

\newcommand{\interior}[1]{\accentset{\smash{\raisebox{-0.12ex}{$\scriptstyle\circ$}}}{#1}\rule{0pt}{2.3ex}}
\definecolor{DGREEN}{rgb}{0,0.65,0.65}
\definecolor{grey1}{rgb}{0.52, 0.52, 0.51}

\usepackage[normalem]{ulem}
\usepackage{color}
\definecolor{blue}{rgb}{0,0,1}
\definecolor{red}{rgb}{1,0,0}

\makeatletter
\newcommand*{\bigcdot}{}% Check if undefined
\DeclareRobustCommand*{\bigcdot}{%
	\mathbin{\mathpalette\bigcdot@{}}%
}
\newcommand*{\bigcdot@scalefactor}{.5}
\newcommand*{\bigcdot@widthfactor}{1.15}
\newcommand*{\bigcdot@}[2]{%
	\sbox0{$#1\vcenter{}$}% math axis
	\sbox2{$#1\cdot\m@th$}%
	\hbox to \bigcdot@widthfactor\wd2{%
		\hfill
		\raise\ht0\hbox{%
			\scalebox{\bigcdot@scalefactor}{%
				\lower\ht0\hbox{$#1\bullet\m@th$}%
			}%
		}%
		\hfil
	}%
}
\makeatother
\DeclareFontFamily{OT1}{rsfs}{} \DeclareFontShape{OT1}{rsfs}{m}{n}{
	<-7> rsfs5 <7-10> rsfs7 <10-> rsfs10}{}
\DeclareMathAlphabet{\mathscr}{OT1}{rsfs}{m}{n}

\def\sc{{\hskip 3.5pt {{}^{{}^{{}_{{}_{\bowtie}}}}} \kern -8.pt{}}}  
\def\SC{{\hskip 3.5pt {{}^{{}^{{}^{{}_{{}_{\bowtie}}}}}} \kern -10.5pt{}}}

\newtheorem{theorem}{Theorem}%[section]
\newtheorem*{theorem*}{Theorem}%[section]
\newtheorem{proposition}{Proposition}%[section]

\newtheorem*{example*}{Example}%[section]
\newtheorem{condition}{Condition}%[section]
\newtheorem{definition}{Definition}%[section]
\usepackage[normalem]{ulem}
\usepackage{color}

\newcounter{mnotecount}%[section]

\newcommand{\mnotex}[1]%{}
{\protect{\stepcounter{mnotecount}}$^{\mbox{\footnotesize $\bullet$\themnotecount}}$ 
	\marginpar{\color{red}
		\raggedright\tiny\em
		$\!\!\!\!\!\!\,\bullet$\themnotecount: #1} }
	
\usepackage{pdfrender}	

\begin{document}

\title{\vskip-1cm\textbf{Spacetime singularities and  curvature blow-ups}%\,\footnote{To appear in the topical volume `Singularity theorems, causality, and all that (SCRI21)' https://link.springer.com/collections/hjjgajaagg} 
}
%{Spacetime singularities and \\ spacetime extensions}

\author{Istv\'an R\'acz \footnote{E-mail address: {\tt racz.istvan@wigner.hu}}}	

\affil{Wigner RCP \\ H-1121 Budapest, Konkoly Thege Mikl\'{o}s \'{u}t  29-33, Hungary}

\maketitle

\begin{abstract}
	The singularity theorems of Penrose, Hawking, and Geroch predict the existence of incomplete inextendible causal geodesics in a wide range of physically adequate spacetimes modeling the gravitational collapse of stars and the expanding universe. Here, using results on spacetime extensions, it is shown that if a suitable low regular form of the strong cosmic censor hypothesis holds, then parallelly propagated blow-up of either the tidal force or frame-drag part of the curvature must occur in ``generic'' timelike geodesically incomplete maximal Cauchy developments. 		
\end{abstract}

%"The whole problem with the world is that fools and fanatics are always so certain of themselves, but wiser men so full of doubts" Bertrand Russell

%%%%%%%%%%%%%%%%%%%% INTRODUCTION %%%%%%%%%%%%%%%%%%%%%%%%%%%%%%%%%%%%%%

\section{Introduction}\label{introduction}
\setcounter{equation}{0} 	

During the first five decades of Einstein's theory of gravity, singular behavior popped up in many of the physically relevant exact solutions. Notably, in these spacetimes, ``singularities'' were always signified by unbounded curvature. Nevertheless, the general covariance of the theory made even the determination of spacetime-singularity to be one of the most intriguing and challenging issues in general relativity \cite{geroch}. In the 1960s, this yielded a fierce debate concerning the relevance of spacetime singularities found in models admitting symmetries \cite{L-K, D-Z-N}. In 1965, by applying methods of global differential geometry, Roger Penrose proved that singularities must occur irrespective of symmetries in spacetimes modeling gravitational collapse \cite{p1,p2}. More precisely, Penrose, in his seminal work, demonstrated that a spacetime cannot be null geodesically complete if the following three conditions are satisfied: 
\begin{itemize}

	\setlength\itemsep{-0.1em}  
	
	\item[(1)] the null convergence condition holds, i.e.,
	$R_{ab}k^ak^b\geq 0$ for all null vectors $k^a$\,,
	
	\item[(2)] the spacetime is globally hyperbolic with a
	non-compact Cauchy surface $\Sigma$\,,
	
	\item[(3)] there exists a closed trapped surface $\mathscr{T}$.  
	
\end{itemize}
The proof of this theorem is derived by contradiction (see, e.g., \cite{Senovilla-1998, Senovilla-2021}). It starts by assuming that the spacetime is null geodesically complete. Then, condition (3) shows that these geodesics begin to focus, whereas condition (1) guarantees that they keep focusing, and focal points must develop. As opposed to this, if the spacetime is null geodesically complete, condition (2) can be used to exclude the appearance of such focal points. The contradiction is avoided by dropping the indirect assumption, verifying that the spacetime cannot be null geodesically complete.

\medskip

It formulates the expectation that gravitational interaction is attractive because null geodesic congruences inevitably get focused by the curvature. The first half of condition (2) is also moderate, as Einstein's theory of gravity is known to possess a well-posed initial value problem.

Note also that condition (1) is very mild. It formulates the expectation that gravitational interaction is attractive in the sense that, if it holds, null geodesic congruences inevitably get focused by the curvature. The first half of condition (2) is also moderate as Einstein's theory of gravity is known to possess a well-posed initial value problem \cite{Choquet, Christodoulou, wald}. The second half assumes only that the spacetime represents the history of an isolated object. Assumption (3) is more demanding. It presumes that the gravitational field is so strong in a domain bounded by a two-dimensional spatial surface $\mathscr{T}$ that the outward-directed null rays, starting at $\mathscr{T}$ perpendicularly, have already negative expansion. This latter condition was verified to hold if a sufficient amount of energy/matter is concentrated in the spatial region bounded by $\mathscr{T}$ \cite{Schoen-Yau, Clark}. 

\medskip

The strength of this theorem is rooted in that all the three conditions above are of pure geometrical character. The same assumptions could be imposed in a wide range of metric theories of gravity. Notably, by adopting the new technical elements applied in Penrose's theorem, Hawking immediately proved the existence of incomplete causal geodesic curves in cosmological models \cite{h}. Soon after, a series of novel singularity theorems by Penrose, Hawking, and Geroch concluded that gravitational singularities occur in many physically realistic situations \cite{h, hp, he}. Note, however, that all these theorems share a shortcoming. Namely, the existence of incomplete, inextendible causal geodesics is used as a synonym of spacetime-singularity \cite{geroch}. 

\medskip

In the following decades considerable efforts had been made to over-bridge the gap between physical intuition and the conclusion of the singularity theorems. All these aimed to verify that some physically relevant quantities do indeed become infinite along the incomplete inextendible causal geodesics predicted by the singularity theorems. Nevertheless, still, no such satisfactory reasoning exists yet. Our main goal in this paper is to bring the intuitive picture of singularities and the predictions of the singularity theorems closer to each other.

\medskip 

While making the scope of the discussions a bit wider, recall first that in metric theories of gravity, such as Einstein’s theory, a spacetime, $(M,g_{ab})$ is supposed to be a smooth Hausdorff, paracompact, connected, orientable manifold $M$. It is also assumed that a smooth metric $g_{ab}$ of Lorentzian signature is also given on $M$ \cite{he, wald}. The base manifold $M$ is also assumed to be chosen sufficiently large to represent all the events compatible with the history of the investigated physical system. In contrast, the presence of incomplete inextendible causal geodesics ---the singularity theorems predicted these--- might be considered as a warning signal indicating that some parts of those spacetimes which describe the expanding universe and the gravitational collapse of stars are missing \cite{geroch}.

\medskip 

After getting aware that spacetimes may not be complete, it is natural to ask if they can be extended. In answering this question, recall first that the above determination of a spacetime refers merely to its essential mathematical structures. In many cases, mainly for mathematical conveniences, it is assumed that a spacetime possesses a smooth differentiable structure and the other fields are also smooth. Recall that in the smooth setting, a seminal result by Choquet-Bruhat and Geroch \cite{Choquet-Geroch} guarantees the existence of a unique (up to diffeomorphisms) maximal Cauchy development (see also \cite{Sbierski-1}). Note also that recently the existence and uniqueness of maximal global hyperbolic developments of vacuum general relativistic initial data sets $(h,K)$ in Sobolev spaces $H^s\oplus H^{s-1}$ was proven in \cite{Chrusciel}, for $s\in \mathbb{N}$ with $s> n/2+1$, where $n\;(\geq 3)$ stands for the dimension of the initial data surface. Given these results, at least in the case of globally hyperbolic spacetimes, we would expect that it is entirely satisfactory to work with the maximal Cauchy development. 

\medskip 

Note, however, that things are much more intricate. Namely, causal geodesically incomplete spacetimes exist such that they contain as a part “the maximal” Cauchy development of suitable smooth initial data given on an otherwise also “maximal” initial data surface. These Cauchy developments can be continued beyond the Cauchy horizon such that no curvature blow-up occurs while crossing this horizon.
As immediate examples, think of the maximal analytic extensions of the Kerr and Taub-NUT spacetimes (see, e.g., \cite{he}). In both cases, the predictive power of general relativity is lost while crossing the Cauchy horizon of the maximal Cauchy development. It is also fair noting that this type of behavior has been found to occur exclusively in spacetimes with symmetries.

\medskip 

The strong version of the cosmic censorship conjecture of Penrose emerged from these troublesome circumstances \cite{Penrose-CCC1, Penrose-CCC2}. Penrose’s strong cosmic censorship conjecture claims that the maximal Cauchy development of a “generic” compact or asymptotically flat initial data is never part of a larger spacetime \cite{Penrose-CCC2}. If true, the corresponding Cauchy development cannot have a Cauchy horizon, especially no extension beyond it could make sense. 
While investigating issues related to the strong cosmic censor hypothesis and the main dilemmas on spacetime singularities, the following questions arose: How could it be shown that something violent happens “there”. Where and what sort? (For detailed discussions on many fundamental issues, see, e.g., \cite{he, Penrose-techniques, geroch, wald,Senovilla-1998,Senovilla-2021}.) In attempting to answer some of these questions, it is rewarding to have a glance at the main cornerstones we already have in our hands. These are the singularity theorems, the existence of maximal Cauchy development, and we may also adopt the strong cosmic censorship conjecture.

\medskip 

Inspecting these fundamental concepts for some time, the following argument, based on contradiction, develops: Consider a causal geodesically incomplete spacetime. Assume that it is the maximal Cauchy development of some “generic” compact or asymptotically flat initial data and that the strong cosmic censor conjecture holds for this class of spacetimes. These assumptions guarantee that such spacetime cannot be extended within the considered differentiability class. Assume, in addition, that nothing violent happens along either of the incomplete inextendible causal geodesics. In particular, assume that all the tidal-force and frame-drag parts of the curvature tensor remain bounded while approaching the “ideal endpoints” of the incomplete inextendible causal geodesics. This regularity of the curvature permits a global extension of the otherwise maximal Cauchy development. If the strong cosmic censor hypothesis holds, this contradiction allows us to conclude that the causal geodesic incompleteness in a maximal Cauchy development must always be accompanied by the singular behavior of some of the tidal-force and frame-drag parts of the curvature tensor.

\medskip 

In this paper, attention will be restricted to the timelike case. In other words, spacetimes admitting incomplete inextendible timelike geodesics will be considered. One of the main results of this paper can be summarized as follows: Consider an $n$-dimensional smooth ``generic'' globally hyperbolic spacetime $(M,g_{ab})$ and assume that $\gamma$ is an incomplete timelike geodesic that is inextendible in $(M,g_{ab})$. Assume also that there exists an $(n-1)$-parameter congruence of causal geodesics, $\mathcal{G}$, spanning an open neighborhood of a final segment of $\gamma$, such that the tidal force and frame-drag (or the electric and magnetic)\,\footnote{For the definition of the tidal force and frame-drag (or the electric and magnetic) parts of curvature see equation \eqref{eq: tidal-frame-drag} in subsection \ref{subsec: gaussian coordinates} below.} parts of the curvature tensor, along with the line integral of the  first-order transversal covariant derivatives of the tidal (or electric) part, ---measured with respect to a parallelly propagated synchronized basis field--- are guaranteed to be uniformly bounded along the members of $\mathcal{G}$. Then $(M,g_{ab})$ is extendible within the class of $C^{0}$ Geroch-Traschen regularity class.

\medskip 

The paper is organized as follow: In Section \ref{Sec: ext-func}  basic results concerning the extendibility of real function, defined on bounded subsets of $\mathbb{R}^n$, is considered. In Section \ref{sec: spa-ext} the basic notion of spacetime extensions and some of the relevant results are recalled. A specific choice for the differentiability class of metric is made and the low differentiable version of the strong cosmic censor hypothesis is discussed in Section \ref{sec: diff-class}. The main results of the present paper are presented in Section \ref{sec: main}, while our conclusions and the final remarks are given in Section \ref{sec: final}.

\section{Extensions of real functions}\label{Sec: ext-func}

In demonstrating that a spacetime is part of a larger one, we need results on the extendibility of spacetimes. Nevertheless, as a preparation, it is rewarding to glance at some of the essential ingredients of this notion. 

\medskip

Recall first that spacetimes are represented by $n$-dimensional smooth differentiable manifolds on which suitably regular metrics are also given. Note that as a manifold locally is $\mathbb{R}^n$, and the components of the metric in the corresponding local coordinates are real functions, it is advantageous to know if real functions, given on a subset of $\mathbb{R}^n$, can be extended. Exactly this problem was studied by Whitney in the early 30's \cite{w1,w2} (see also \cite{Fefferman}). He considered a real-valued function $\mathcal{F}$, say of class $C^m$, defined on a subset $\mathscr{A}$ of $\mathbb{R}^n$, and asked under which conditions exists a function $\widetilde{\mathcal{F}}$, of class $C^\ell$, with $\ell\leq m$, on the entire of $\mathbb{R}^n$, such that $\widetilde{\mathcal{F}}= \mathcal{F}$ on $\mathscr{A}$? 

\medskip

In answering the above addressed issues Whitney in \cite{w1} introduced the term ``property $\mathscr{P}$'', to characterize subsets in $\mathbb{R}^n$, defined as follows.\footnote{In the literature, property $\mathscr{P}$ is also often termed quasi-convex.}

\begin{definition} 
		A point set $\mathscr{A}\subset\mathbb{R}^n$  is said
		to possess the property $\mathscr{P}$ if there is a positive real number $\omega$ such that for any two points $x$
		and $y$ of $\mathscr{A}$ can be joined by a curve in $\mathscr{A}$ of length ${L}\leq \omega\cdot \rho(x,y)$, where $\rho(x,y)$ denotes the Euclidean distance of the points $x, y \in\mathbb{R}^n$. 
\end{definition}

The main result by Whitney can be summarized by the following: 

\begin{theorem}\label{theor: wt}
	Assume that $\mathscr{A}\subset \mathbb{R}^n$ has property $\mathscr{P}$, and
	that $\mathcal{F}(x^1,...,x^n)$ is of class $C^m$, for some positive integer 
	$m\in \mathbb{N}$, in $\mathscr{A}$. Suppose that $\ell\in \mathbb{N}$
	is so that $\ell\leq m$, and also that each of the $\ell^{th}$ order derivatives
	$\partial_{x^1}^{\ell_1}\cdots \partial_{x^n}^{\ell_n} \mathcal{F}$, with
	$\ell_1+\cdots  +\ell_n= \ell$, can be defined on the boundary
	$\partial\mathscr{A}$ of $\mathscr{A}$ so that they are continuous in
	$\overline{\mathscr{A}}=\mathscr{A}\cup\partial\mathscr{A}$. Then there
	exists an extension $\widetilde{\mathcal{F}}$ of $\mathcal{F}$ so that
	$\widetilde{\mathcal{F}}$ is of class $C^\ell$ throughout
	$\mathbb{R}^n$. 
\end{theorem}

Note that it was shown in \cite{w1,w2} that $\widetilde{\mathcal{F}}$ can be chosen smooth (or, if needed, it can be analytic) in $\mathbb{R}^n\setminus\overline{\mathscr{A}}$. 

To demonstrate that property $\mathscr{P}$ plays an essential role in Theorem \ref{theor: wt}, it is illuminating to recall Example 4.1.\;of \cite{racz-2}.  A smooth real function $\mathcal{F}$ on a bounded subset $\mathscr{A}\subset\mathbb{R}^2$ is constructed there such that $\mathcal{F}$, along with its partial derivatives up to any fixed order, are uniformly bounded in $\mathscr{A}$, nevertheless, $\mathcal{F}$ cannot even have a continuous extension to the closure of $\mathscr{A}$. 

Note, finally, that the property $\mathscr{P}$ as a concept has nothing to do with the causal structure of an underlying spacetime, even if it is applied to characterize coordinate patches therein. Nevertheless, it is worth emphasizing that in the case of globally hyperbolic spacetimes ---these are at the center of the investigations in this paper--- the coordinate domains applied in our constructions, in Section \ref{sec: main}, are guaranteed to possess the property $\mathscr{P}$ (see Proposition 4.1.\;in \cite{racz-2}).

\section{Spacetime extensions}\label{sec: spa-ext}

Note first that while giving the notion of spacetime extensions, various differentiability assumptions on the metric have been applied depending on the context. In addition, as we have not yet fixed a preferred differentiability class, for the moment, it is advantageous to keep the applied differentiability class as flexible as it is possible. Accordingly, in the definition below $C^{X}$ will signify either the class of analytic, $C^\infty$, $C^{k}$, $C^{k-}$, $C^{k,\alpha}$ functions. It may also stand for more involved classes such as the differentiability class $\mathcal{C}^{0-,\alpha}$ applied in \cite{c3}. With this notation, spacetime extensions can be defined as follows:

\begin{definition}
Let $M$ and $\widehat{M}$ be $n$-dimensional connected, paracompact, Hausdorff, smooth differentiable manifolds, and $(M,g_{ab})$ and $(\widehat{M},\widehat{g}_{ab})$ be time oriented spacetimes with metrics at least of class $C^{X}$. Then  $(\widehat{M},\widehat{g}_{ab})$ is called to be a $C^{X}$-extension of $(M,g_{ab})$ if there exists an embedding $\Phi: M \rightarrow \widehat{M}$  such that $\Phi[M]$ is a proper subset of $\widehat{M}$ and $\Phi$ is a diffeomorphism between $M$ and $\Phi[M]\subset \widehat{M}$, and such that $\Phi$ carries the metric $g_{ab}$ into $\widehat{g}_{ab}|_{\Phi[M]}$, i.e., $\Phi^*g_{ab}=\widehat{g}_{ab}|_{\Phi[M]}$. If $(M,g_{ab})$ admits a $C^{X}$-extension it is said to be $C^{X}$-extendible. If such an extension does not exist $(M,g_{ab})$ is called to be $C^{X}$-inextendible.
\end{definition}

%\bigskip

%\centerline{\includegraphics[width=9cm]{../TEX/eload/2021/EREP-2021/abrak/ext.pdf}}

%\bigskip

Note that the involved manifolds cannot be rougher than $C^1$ to permit (at least)  continuous tangent spaces and also to be able to host a continuous metric.  Nevertheless, as it is argued in Theorem 2.9 in \cite{hirsch} if a $C^r$-differentiability structure, $r\geq 1$ is given on $M$, then for every $s$, $r < s \leq \infty$ there exists a compatible  $C^s$-differentiability structure on $M$ such that it is unique up to $C^s$-diffeomorphisms, and such that it is $C^r$-diffeomorphic to the original one. Therefore, without loss of generality, we assumed above that both of the manifolds, $M$ and $\widehat{M}$, admit smooth differentiable structure. 
Note also that the differentiability class of $g_{ab}$ need not to be exactly $C^{X}$, i.e.,  $g_{ab}$ may belong to some higher differentiability class. 

\medskip

We close this section by briefly recalling some of the most important results on spacetime extensions. In doing so, note that the first systematic investigation of spacetime extensions was carried out by Clarke \cite{c1,c2,c3,c4}. His main result is that for a ``generic'' globally hyperbolic $\mathcal{C}^{0-}$ causal geodesically incomplete spacetime, there is a $\mathcal{C}^{0-,\alpha}$ extension provided that the Riemann tensor is also H\"older-continuous. (A spacetime, in \cite{c1}\cite{c3}, was considered generic if its $b$-completion was not $D$-specialized at any of the $b$-boundary points attached to it to represent singularities.) Besides the indisputable importance of these pioneering investigations, there are some drawbacks to the above-recalled result. Firstly, Clarke's results are based on an extensive use of the $b$-boundary construction, which is known to have severe defects even for the simplest  Friedman-Robertson-Walker cosmological model (for more details, see section 5.2 of \cite{c3}). Secondly, it may happen that a given spacetime cannot be extended within the $\mathcal{C}^{0-,\alpha}$ class; in contrast that the curvature remains finite everywhere, simply because it fails to be H\"older-continuous at the points of the $b$-boundary.

\medskip

Given the indicated drawbacks, it became of obvious interest to construct spacetime extensions using the regular geometrical structures of the spacetime to be extended exclusively. In particular, it is preferable to avoid using boundary constructions. Keeping these ideas in the forefront, the present author also conducted a systematic study of local and global extensions of causal geodesically incomplete spacetimes \cite{racz-1, racz-2}. The main result in \cite{racz-2} can be summarized as follows: Consider an $n$-dimensional smooth ``generic'' (i.e., locally algebraically non-special) globally hyperbolic spacetime $(M,g_{ab})$ and assume that $\gamma$ is an incomplete causal geodesic that is inextendible in $(M,g_{ab})$. Assume also that there is an $(n-1)$-parameter congruence of causal geodesics, $\mathcal{G}$, spanning an open neighborhood of a final segment of $\gamma$, such that the components of the curvature tensor, along with its covariant derivatives up to order $(k-1)$, and also the line integrals of the components of the $k$th-order covariant
derivatives are finite along the members of $\mathcal{G}$ ---stipulated with respect to a parallelly propagated synchronized basis field--- are guaranteed to be uniformly bounded along the members of $\mathcal{G}$. Then $(M,g_{ab})$ is $C^{k-}$-extendible. Comparing these results with those covered by the present paper, it is transparent that much lower differentiability requirements suffice to show the extendibility of smooth global hyperbolic spacetimes within the class of continuous Geroch-Traschen metrics. Note also that the restrictions on curvature are more optimized here as, on the way of proving our main results (see, i.e., Theorems \ref{theor: main} and \ref{theor: glob-ext} below), we refer merely to the tidal force and frame-drag parts of the curvature.

\section{The choice of differentiability class}\label{sec: diff-class}

In proceeding, we now fix the differentiability class of the metric that suits most of the following discussions. In doing so, recall first that general relativity is a physical theory; thereby, the field equations and their solvability or the possible breakdown of the field equations are of fundamental interest to us. On these grounds, it is desirable to admit spacetime models with metrics and other fields that are less well behaved than smooth. It is also apparent that the wider the differentiability class of involved metrics and matter fields is, the wider the class of gravity-matter systems that can be studied within the selected framework.

\medskip

These observations immediately suggest involving the widest class of spacetime models, allowing us to make sense of the Einstein equations at least as distributions. Geroch and Traschen investigated this fundamental issue in \cite{Geroch-Traschen}. They showed that the widest possible class of metrics for which the Riemann, Einstein, and Weyl tensors make sense as distributions is the space of ``regular metrics'' or, as we also refer to it, the ``Geroch-Traschen regular metrics''. In particular, $g_{ab}$ is said to be a Geroch-Traschen regular metric if

\begin{itemize}
		
	\setlength\itemsep{-0.1em}
		
	\item[(1)] $g_{ab}$ locally bounded with a locally bounded inverse $g^{ab}$,\,\footnote{Here, as in \cite{Geroch-Traschen}, to have a well-defined inverse, it is tacitly assumed that $g_{ab}$ is non-degenerate. In the present context, this can be guaranteed if, on compact sets, a lower uniform bound exists on the determinant \cite{LeFloch-Madare}.}
	
	\item[(2)] the ``weak derivatives'' of $g_{ab}$ are locally square-integrable.
		
\end{itemize}

\medskip

The metric $g_{ab}$ and its inverse $g^{ab}$ are locally bounded if the scalar densities $g_{ab}t^{ab}$ and $g^{ab}u_{ab}$ are bounded for all test fields $t^{ab}$ and $u_{ab}$. A test field is supposed to be a smooth tensor density of compact support. The tensor distributions are continuous linear maps from the vector space of test fields to the real numbers. The derivative  of the locally bounded metric $g_{ab}$, as distribution, is the distribution $\overline{\nabla}_eg_{ab}$ such that
\begin{equation}\label{eq: der-of-distr}	
	\int_{M}(\overline{\nabla}_e g_{ab})t^{eab} = -\int_{M}g_{ab} (\overline{\nabla}_et^{eab})	
\end{equation}
for any test field $t^{abc}$, where  $\overline{\nabla}_a$ is a smooth torsion-free covariant derivative operator on $M$. Note that the integrals in \eqref{eq: der-of-distr} make sense as test fields are assumed to be tensor densities of weight  $-1$ \cite{Geroch-Traschen}. Then a locally integrable\,\footnote{A tensor field $\nu_{a\dots b}{}^{c\dots d}$, defined almost everywhere on $M$,  is called locally integrable if for every test field $t^{a\dots b}{}_{c\dots d}$ the scalar density $\nu_{a\dots b}{}^{c\dots d}\,t^{a\dots b}{}_{c\dots d}$ is Lebesgue measurable and its Lebesgue integral converges.} tensor field $\mu_{abc}$ is called the weak-derivative of $g_{ab}$ if 
\begin{equation}	
	\int_{M} g_{ab}(\overline{\nabla}_et^{eab}) = -\int_{M}\mu_{eab}\,t^{eab} 	
\end{equation}
holds for all test fields $t^{abc}$. The weak-derivative $\mu_{abc}$ of $g_{ab}$ is called locally square integrable if the scalar density $\mu_{abc}\,\mu_{def}\,t^{abcdef}$ is locally integrable for all test fields $t^{abcdef}$.

\medskip

It was proved in \cite{Geroch-Traschen} that if the metric $g_{ab}$ is Geroch-Traschen regular, then the Riemann, Einstein, and Weyl tensors make sense as distributions. Geroch and Traschen also showed that if a regular metric, $g_{ab}$, is not only locally bounded with a locally bounded inverse but is continuous, then it can be approximated by sequences of smooth metrics $\{{}^{(i)}{}g_{ab}\}$ so that the associated curvature tensors $\{{}^{(i)}R_{abc}{}^d\}$ converge in $L^2$ to the curvature distribution assigned to the continuous regular metric $g_{ab}$. 

\medskip

Returning to our main issue, given the results in \cite{Geroch-Traschen} it is tempting to choose the $C^{0}$ Geroch-Traschen regularity class. This choice is also supported by the fact that it is considered the broadest class of metrics such that weak solutions of the Einstein equations exist (see, e.g., \cite{he, Geroch-Traschen, Christodoulou, Dafermos}).
It is also frequently claimed (see, e.g., \cite{Christodoulou, Dafermos}) that local existence and uniqueness of weak solutions to the initial value problem of the vacuum Einstein's equations are expected to exist in a setting with metrics that belong to the $C^{0}$ Geroch-Traschen regularity class. Note also that regardless of the differentiability class, it is an additional requirement that maximal Cauchy developments should also exist. If one can extend a maximal Cauchy development of the data given on a maximal initial data slice, the extension is inherently ambiguous beyond the Cauchy horizon since, even if the vacuum Einstein equations are satisfied there, the data on the initial data surface no longer determine the solution beyond the Cauchy horizon. To over-bridge the gap between available techniques and firm results, we shall adopt the following $C^0$ form of the strong cosmic censor conjecture proposed by Chrishtoduolus, Dafermos, Sbierski \cite{Christodoulou, Dafermos, Sbierski-2}:  

\begin{condition}\label{cond: SCCH}
	The maximal globally hyperbolic development of generic compact or asymptotically flat initial data is inextendible as a spacetime within the class of continuous Geroch-Traschen regular metrics.	
\end{condition}

While it appears conceivable that  Condition \ref{cond: SCCH} holds, this remains to be seen. As there are a lot of missing technical elements yet it is rewarding to have a glance of the regularity class of $C^{0,1}_{loc}$ metrics. In doing so note first that the Geroch-Traschen regular metrics belong to the intersection $H^1_{loc}\cap L^\infty_{loc}$, where the relation $H^1 = W^{1,2}$, valid for the $L^2$-based Sobolev spaces, was used \cite{af}. Taking then into account that on any domain $C^{0,1}_{loc}= W^{1,\infty}_{loc}$, and that the local Sobolev spaces are nested, i.e., $W^{1,m}_{loc} \subseteq W^{1,n}_{loc}$ for any $m \geq n$ hold, it follows that $C^{0,1}_{loc}$ metrics belong to the Geroch-Traschen regularity class and as such they possess a distributional curvature and Einstein tensor. Having this on mind, it is also worth mentioning that several recent results holds for the regularity class of $C^{0,1}_{loc}$ metrics.  
For instance, it was shown in \cite{Chrusciel-Grant} that $C^{0,1}_{loc}$ differentiability of the metric suffices for many vital results of the $C^\infty$ causality theory. Analogously, within the $C^{0,1}_{loc}$  regularity class global hyperbolicity makes sense, allowing to represent Cauchy hypersurfaces as level sets of $C^\infty$ time functions \cite{Chrusciel, Samann}. It is also worth noting that the global existence and uniqueness of solutions to linear wave equations can also be guaranteed if the background metric belongs to the $C^{0,1}_{loc}$ regularity class \cite{Chrusciel, sanchez-vickers}. In this context, it is also worth mentioning that the existence of maximal Cauchy development for the ``$3+1$'' vacuum Einstein equations requires a metric with critical Sobolev exponent $s=2$ \cite{Klainerman-Rodnianski}.     

\medskip

Finally, it is also worth emphasizing that many physically adequate solutions belong to the class of spacetimes with $C^0$ Geroch-Traschen regular metric. They include, for instance, gravitational shock waves (where the curvature has a $\delta$-function behavior on a null three-surface) \cite{Penrose-pp}, thin mass shells (where the curvature has a $\delta$-function behavior on a timelike three-surface) \cite{Israel}. Analogously, this class also involves solutions containing pressure-free matter where the geodesic flow lines have two- or three-dimensional caustics and shell-crossing singularities \cite{Szekeres}. These examples demonstrate that whenever Condition 1.\;holds the existence of incomplete timelike geodesics, within the class of spacetimes with  $C^0$ Geroch-Traschen regular metrics, cannot simply be explained by referring to the presence of certain non-smooth wavefronts or caustics of flow lines. Instead, they indicate more serious breakdowns of physics. Indeed, our principal aim in the present paper is to show that given a ``generic'' (i.e., locally algebraically non-special) timelike geodesically incomplete globally hyperbolic spacetime with a smooth metric, either the spacetime can be extended within the class of $C^0$ Geroch-Traschen regular metrics or some of the tidal force or frame-drag  parts of the curvature tensor, or the  first-order transversal covariant derivatives of the tidal forces ---measured by a $(n-1)$-parameter family of synchronized observers\,\footnote{In Subsection \ref{subsec: gaussian coordinates} below,  these type of observers are modeled by applying $(n-1)$-parameter families of synchronized timelike geodesics and frame fields parallelly propagated along them.} --- become unbounded.

\section{The main results}\label{sec: main}

This section is to present our main results. The first one read as: 
\begin{theorem}\label{theor: main}
	Consider an $n$-dimensional smooth locally algebraically non-special maximal globally hyperbolic timelike geodesically incomplete spacetime. Assume that Condition\,\ref{cond: SCCH}  is satisfied, i.e., the $C^0$ form of the strong cosmic censor hypothesis holds. Denote by $\gamma$ one of the incomplete timelike geodesics, and consider a synchronized $(n-1)$-parameter family of timelike geodesics, $\mathcal{G}$, spanning a neighborhood of a final segment of  $\gamma$. Then, in any arbitrarily small neighborhood of $\gamma$, there exists a member $\overline{\gamma}$ of $\mathcal{G}$ such that either one of the tidal force or frame-drag components of the curvature or one of the first-order transversal covariant derivatives of one of the tidal force components blows up along $\overline{\gamma}$.
\end{theorem}

{\noindent \bf Proof:} The proof of this theorem is based on contradiction. It starts by assuming that the tidal force and frame-drag parts of the curvature, along with the  first-order transversal covariant derivatives of the tidal forces, ---measured with respect to a synchronized basis field defined along the $(n-1)$-parameter family of synchronized timelike geodesics,  $\mathcal{G}$,--- remain uniformly bounded along the members of $\mathcal{G}$. This allows proving Theorem \ref{theor: glob-ext} below claiming that the spacetime can be extended within the $C^0$ Geroch-Traschen regularity class. This, however, is incompatible with the assumptions we made. Namely, a ``generic'' smooth maximal globally hyperbolic timelike geodesically incomplete spacetime cannot be extended when the  $C^0$ version of the strong cosmic censorship hypothesis is also assumed to hold. 

Accordingly, proving Theorem \ref{theor: glob-ext} below, given the above reasoning, completes the proof of Theorem \ref{theor: main}. 
{\hfill$\Box$}

\begin{theorem}\label{theor: glob-ext}
	Consider an $n$-dimensional locally algebraically non-special smooth globally hyperbolic timelike geodesically incomplete spacetime $(M,g_{ab})$. Denote by $\gamma$ one of the incomplete timelike geodesics, and consider a synchronized $(n-1)$-parameter family of timelike geodesics, $\mathcal{G}$, spanning a neighborhood of a final segment of  $\gamma$. Assume that the tidal force and frame-drag parts of the curvature, along with the line integrals of the first-order transversal covariant derivatives of the tidal forces, are uniformly bounded along the members of $\mathcal{G}$. Then $(M,g_{ab})$  can globally be extended within the $C^0$ Geroch-Traschen regularity class. 
\end{theorem}
{\noindent \bf Proof:} The proof of this theorem is given by performing the following sequence of steps. 
\begin{enumerate}
	
	\setlength\itemsep{-0.1em}
	
	\item  Assume that $\gamma: (t_1,t_2)\rightarrow M$ is a future directed and future incomplete timelike geodesic that is inextendible in $(M,g_{ab})$. For definiteness
	we assume that $\gamma$ is future incomplete, the other case then follows by
	a time reversal.
	\item  Choose an $(n-1)$-parameter family of synchronized future directed timelike geodesics,  $\mathcal{G}$, such that $\gamma$ belongs to $\mathcal{G}$.
	\item  Choose ${\mathcal{U}} \subset M$ such that  ${\mathcal{U}}$ is the union of the images of the members of $\mathcal{G}$, and such that a final segment $\gamma|_{[t_0,t_2)}$, for some $t_0\in (t_1,t_2)$ is contained in ${\mathcal{U}}$. 
	\item Show, under the assumption in our theorem, that there exist $\widetilde{\mathcal{U}} \subset \mathbb{R}^n$ and a smooth embedding $\phi:{\mathcal{U}} \rightarrow  \widetilde{\mathcal{U}} $ such that $\phi\circ\gamma$ is continuously extendible in $\widetilde{\mathcal{U}}$. 
	\item Extend the metric $g_{ab}$ from ${\mathcal{U}}$ to $\widetilde{\mathcal{U}}$ within the  $C^0$ Geroch-Traschen regularity class, and denote by $\phi: (\mathcal{U},g_{ab}\vert_{\mathcal{U}}) \rightarrow (\widetilde{\mathcal{U}},\widetilde{g}_{ab})$ the intermediate extension obtained. 
	\item The desired $\Phi: (M,g_{ab}) \rightarrow (\widehat{M},\widehat{g}_{ab})$ global extension is defined then by  gluing  $(M,g_{ab})$ and $(\widetilde{\mathcal{U}},\widetilde{g}_{ab})$ at their ``common parts'', i.e., applying a quotient space induced by $\phi$, whereas 
	the metric $\widehat{g}_{ab}$ gets to be determined by $g_{ab}$ and $\widetilde{g}_{ab}$.
\end{enumerate}

\subsection{Synchronized Gaussian coordinates and %synchronized 
	reference frames}\label{subsec: gaussian coordinates}

This section is to introduce the synchronized Gaussian coordinates and synchronized orthonormal basis fields that play a central role in constructing the desired intermediate and global extensions.

\medskip

Consider a future directed and future incomplete timelike geodesic $\gamma: (t_1,t_2)\rightarrow M$. Assume that $t$ is the proper time parameter along  $\gamma$, and that $v^a$ is the corresponding unit tangent field $v^a=(\partial/\partial t)^a$ along  $\gamma$.  
Then  synchronized Gaussian coordinates can be defined, in a sufficiently small neighborhood of any point $p=\gamma(t_0)$ of $\gamma$, with  $t_0\in(t_1,t_2)$, as follows \cite{racz-1,racz-2}.
Choose first a sufficiently small open neighborhood, $Q$, of the origin in the linear subspace $T_p^\perp(M)$, spanned by the $(n-1)$-dimensional subspace of spacelike vectors orthogonal to $v^a$, such that the exponential map\footnote{%  
	The exponential map $\exp: T_{p}(M)\rightarrow M$ assigns a point $q\in M$ to a vector $X^a\in T_{p}(M)$ such that $q$ is in unit affine parameter distance from $p$ along the geodesic starting at $p$ with tangent $X^a$.} 
is guaranteed to be a local diffeomorphism between $Q$ and $\exp[Q]\subset M$. Denote by $\Sigma$ the image of $Q$ under the action of the exponential map, i.e., $\Sigma=\exp[Q]$. 
Accordingly, $\Sigma$ is generated by spacelike geodesics starting at $p=\gamma(t_0)$ with tangent orthogonal to $v^a$. Extend then $v^a$ from $p=\gamma(t_0)$ to a smooth future directed normalized, $g_{ab}v^av^b=-1$, timelike vector field on $\Sigma$ which is also everywhere normal to $\Sigma$. Chose $(x^1,\dots,x^{n-1})$ to be arbitrary local coordinates on $\Sigma$ and consider the  $(n-1)$-parameter congruence of timelike geodesics,  $\mathcal{G}$, starting at the points of  $\Sigma$ with tangent $v^a$. Since $\Sigma$ and $v^a$ are smooth these geodesics do not intersect in a sufficiently small neighborhood $\mathnormal{V}$ of $\Sigma$. Extend the functions $x^1,\dots,x^{n-1}$ to $\mathnormal{V}$, by keeping their values constant along the geodesics in $\mathcal{G}$, and chose the proper time $t=x^n$  on $\mathnormal{V}$ as the $n^{th}$ coordinate that is synchronized such that $x^n=t_0$ on $\Sigma$.  The functions $(x^1,\dots,x^n)$ give rise to local coordinates on $\mathnormal{V}$. 

%\bigskip

%\centerline{\includegraphics[width=11cm]{../TEX/eload/2021/EREP-2021/abrak/exp.pdf}}

%\bigskip

\medskip

In these synchronized Gaussian coordinates, the
spacetime metric can be seen to take the form \cite{Penrose-techniques,wald}
\begin{equation}\label{let} 
	ds^2=-dt^2+g_{\alpha\beta}\,dx^\alpha dx^\beta \,,
\end{equation} 
where $g_{\alpha\beta}$ is a $(n-1)\times (n-1)$ positive definite matrix the components of which are smooth functions of all the coordinates $(x^1,\dots,x^n)$, and the Greek indices take the values $1,2,\dots,n-1$.

\medskip

It is also worth mentioning here that, by  construction of the Gaussian coordinates, the coordinate basis fields $E^a_{\alpha}=(\partial/\partial {x^\alpha})^a$, with $\alpha=1,2,\dots,n-1$ are Jacobi fields along the $(n-1)$-parameter congruence of timelike geodesics in $\mathcal{G}$. Accordingly, the $E^a_{\alpha}$ coordinate basis fields are subject to the Jacobi equation  
\begin{equation}
	v^e\nabla_e(v^f\nabla_f E^a_{(\alpha)}) = {R_{efg}}^a v^e
	E^f_{(\alpha)} v^g\,, \label{jac}
\end{equation}
where $\nabla_a$ denotes the metric compatible covariant derivative operator.

\medskip

A synchronized basis field, $\{e_{(\mathfrak{a})}^a \}$, along the members of an $n-1$-parameter family of timelike geodesics $\mathcal{G}$, can also be chosen as follows. Start with an orthonormal basis $\{e_{(\mathfrak{a})}^a \}\subset T_p(M)$, with name index $\mathfrak{a}$, taking the values $1,2,\dots,n$, such that $e_{(\mathfrak{n})}^a=v^a$ at $p$. Extend then $\{e_{(\mathfrak{a})}^a \}$ from $p$ by parallelly propagating it first  along the spacelike geodesics generating $\Sigma$. If $e_{(\mathfrak{n})}^a$ happens to be different from the already defined $v^a$ vector field on $\Sigma$ apply the simplest boost transformation point-wise such that $e_{(\mathfrak{n})}^a=v^a$ holds for the yielded smooth\,\footnote{The smoothness of the  basis field $\{e_{(\mathfrak{a})}^a \}$ on $\Sigma$ follows from that of $v^a$, the process of parallel propagation, along the generators of $\Sigma$, and also that of the boost transformations applied point-wise on $\Sigma$.} 
basis field $\{e_{(\mathfrak{a})}^a \}$  on  $\Sigma$. Finally, extend this smooth basis field by parallelly propagating $\{e_{(\mathfrak{a})}^a \}$ from $\Sigma$, along the members of the $(n-1)$-parameter family of timelike geodesic congruence $\mathcal{G}$.  Note also that, by construction, the relation $e_{(\mathfrak{n})}^a=v^a$ holds along the members of $\mathcal{G}$.

\medskip

Utilizing the above defined synchronized basis field, $\{e_{{{(\mathfrak{a})}}}^a \}$, for instance, the tidal force  and frame-drag (or the electric and magnetic) parts of the Riemann tensor, with respect to the observers moving along the members of $\mathcal{G}$ with velocity $v^a$, can be given, see, e.g., \cite{Ellis, Nicolson},  as
\begin{equation}\label{eq: tidal-frame-drag}
	R_{\mathfrak{a} n\mathfrak{b} n}= R_{abcd}\,e_{{{(\mathfrak{a})}}}^a v^b
	e_{{{(\mathfrak{b})}}}^c v^d \ \ \ {\rm and} \ \ \ 
	R_{\mathfrak{a}\mathfrak{b}\mathfrak{c}  n} = R_{abcd}\,e_{{{(\mathfrak{a})}}}^a 
	e_{{{(\mathfrak{b})}}}^b e_{{{(\mathfrak{c})}}}^c v^d\,,
\end{equation}
respectively,
where the indices $\mathfrak{a},\mathfrak{b}$, $\mathfrak{c}$ take the values $1,2,\dots,n-1$. Note that the tidal force and frame-drag parts of curvature, as given in \eqref{eq: tidal-frame-drag}, are related to the electric and magnetic parts of the curvature, defined with respect to the unite timelike vector field $v^a$ \cite{Ellis}\cite{Nicolson}. In the $4$-dimensional case, the electric and magnetic parts are represented by the symmetric tensors $\mathscr{E}_{\mathfrak{a}\mathfrak{b}}$ and $\mathscr{B}_{\mathfrak{a}\mathfrak{b}}$ defined as
\begin{equation}\label{eq: electric-magnetis}
	\mathscr{E}_{\mathfrak{a}\mathfrak{b}}=R_{\mathfrak{a} n\mathfrak{b} n}= R_{abcd}\,e_{{{(\mathfrak{a})}}}^a v^b
			e_{{{(\mathfrak{b})}}}^c v^d \ \ \ {\rm and} \ \ \  
			\mathscr{B}_{\mathfrak{a}\mathfrak{b}}= \tfrac12\,\epsilon_{ef\mathfrak{a}}R^{ef}{}_{\mathfrak{b}  n} = \tfrac12\,\epsilon_{efa} R^{ef}{}_{bc}\, e_{{{(\mathfrak{a})}}}^a 
			e_{{{(\mathfrak{b})}}}^b  v^c \,,
\end{equation}							
respectively, where $\epsilon_{abc}$ stands for the contraction $\epsilon_{abc}=\epsilon_{abce}v^e$ of the $4$-volume element $\epsilon_{abce}$ and the unite timelike vector field $v^a$.

\subsection{The selection of $\mathcal{U}$}\label{subsec: selection}

Since our principal aim is to perform spacetime extensions based on a suitable choice of synchronized Gaussian coordinates, it is of obvious interest to know how far from $\Sigma$ these 
coordinates and the synchronized basis field can be applied. Our aim in this subsection is to answer this question.

\medskip

Start by choosing $\mathscr{U}\subset\mathbb{R}^n$ consisting of those $n$-tuples, $(x^1,\dots,x^n)$, to which there exists a timelike geodesic $\overline\gamma: (\overline t_1,\overline t_2) \rightarrow M$ in $\mathcal{G}$ such that $x^n\in (\overline t_1,\overline t_2)$ and $(x^1,\dots,x^{n-1},t_0)$ are the coordinates of the intersection $\overline\gamma\cap \Sigma$. Define the map  $\psi:\mathscr{U} \rightarrow M$ such that $\psi(x^1,\dots,x^n)=\overline\gamma(x^n)$ with assuming that $(x^1,\dots,x^{n-1},t_0)$ are the coordinates of $\overline\gamma\cap \Sigma$. Although, $\psi$, by construction, is smooth, in general, it is not necessarily one-to-one. If it is not  one-to-one $\mathnormal{V}$ is a proper subset of $\psi[\mathscr{U}]$.

\medskip

Recall also that the choice  we have made for $\Sigma$ guarantees that the second fundamental form $\chi_{ab}$ of $\Sigma$ vanishes at $p=\gamma(t_0)$. Therefore, by choosing a sufficiently small open
neighborhood $\sigma_{t_0}$ of $p$ with compact closure in $\Sigma$ it can be guaranteed that there exists a small positive number $\varkappa$, depending on $\chi_{ab}$ and  $\sigma_{t_0}$, such that for a suitable norm ---for its definition see Section 3.\;in \cite{racz-2}--- $\|\chi_{ab}\|<\varkappa$ holds on $\sigma_{t_0}$. 

\medskip

Our aim is to show that, in suitable circumstances, there exist $t_0$, $\sigma_{t_0}\subset \Sigma$ and $\varepsilon>0$ such that on the subset $\mathscr{U}_{[\sigma_{t_0},\,\varepsilon]}$ of $\mathscr{U}\subset\mathbb{R}^n$, defined as
\begin{equation}\label{usze}
\mathscr{U}_{[\sigma_{t_0},\,\varepsilon]}:=\{(x^1,\dots,x^{n})
	\in\mathscr{U}\,|\, \psi(x^1,\dots,x^{n-1},t_0)\in \sigma_{t_0}\
	{\rm and}\ x^n\in [t_0,t_2+\varepsilon)\}\,,
\end{equation} 
the map $\psi: \mathscr{U}_{[\sigma_{t_0},\,\varepsilon]} \rightarrow M$ is one-to-one. 
In verifying this claim first we refer to the proof of Proposition 3.2.5 of \cite{racz-2} to the individual members $\overline\gamma \in\mathcal{G}$, the boundedness of the tidal force (or electric) part of the curvature tensor implies that there exist $\epsilon>0$ such that no conjugate point along $\overline\gamma$, in the parameter interval $[t_0,t_2+\varepsilon)$, can occur to $\sigma_{t_0}$.
This, with reference to the claims in the proof of Proposition 3.1 of \cite{racz-2}, implies that to any point $q$ in $\mathscr{U}_{[\sigma_{t_0},\,\varepsilon]}$ there must exist an open neighborhood $\mathcal{O}_{q}$ such that $\psi$ is a local diffeomorphism between $\mathscr{O}_q$ and its image $\mathcal{O}_q=\psi[\mathscr{O}_q]$. In other words, this guarantees that the Gaussian coordinates are locally well-defined on $\mathcal{O}_{q}$.

\medskip

The goal is now to show that under suitable conditions the Gaussian coordinates are only locally well-defined but also globally well-defined throughout  $\psi[\mathscr{U}_{[\sigma_{t_0},\,\varepsilon]}]$. To understand the difficulties at this point recall that spacetimes with ``quasi-regular'' singularities do also exist (see, e.g., Refs.\;\cite{c1,es,es2}) which can get in the way of getting well-defined Gaussian coordinates on $\psi[\mathscr{U}_{[\sigma_{t_0},\,\varepsilon]}]$. These spacetimes are known to have topological defects which prevent the existence of $\mathscr{U}_{[\sigma_{t_0},\,\varepsilon]} \subset \mathbb{R}^n$ on which $\psi$ could be one-to-one. To separate these cases in \cite{racz-2} the notion of topological singularity was introduced which, in the timelike case, reads as: 
\begin{definition}\label{DefTop}	
	A future directed incomplete future inextendible timelike geodesic $\gamma:(t_1,t_2)\rightarrow M$ is said to terminate on a {\it topological singularity} if there is no choice for $t_0$, $\sigma_{t_0}\subset \Sigma$ and $\varepsilon$ such that the set $\psi[\mathscr{U}_{[\sigma_{t_0},\,\varepsilon]}]$ would be simply connected. 	
\end{definition}
It is proved then in Section 5 of \cite{racz-2} that the existence of a topological singularity in globally hyperbolic spacetimes implies that they are locally algebraically special, i.e., these spacetimes cannot be ``generic''. 
It is also proved (see Theorem 3.1.\;in \cite{racz-2}) that whenever $\gamma:(t_1,t_2)\rightarrow M$ does not terminate on a topological singularity for a suitable choice of $\mathscr{U}_{[\sigma_{t_0},\,\varepsilon]}$ the members of $\mathcal{G}$ do not intersect in $\psi[\mathscr{U}_{[\sigma_{t_0},\,\varepsilon]}]$, i.e., there exist $t_0$, $\sigma_{t_0}\subset \Sigma$ and $\varepsilon>0$ such that $\psi$ is one-to-one on the entire of $\mathscr{U}_{[\sigma_{t_0},\,\varepsilon]}$.

\medskip

Choose $\mathcal{U}$ to be the interior, $(\psi[\mathscr{U}_{[\sigma_{t_0},\,\varepsilon]}])\hskip-0.5mm\interior{\phantom{a}}$, of the image $\psi[\mathscr{U}_{[\sigma_{t_0},\,\varepsilon]}]$ of $\mathscr{U}_{[\sigma_{t_0},\,\varepsilon]}$, and also $\widetilde{\mathcal{U}}$ to be the Cartesian product $\varsigma_{t_0}\times[t_0,t_2+\varepsilon) \subset \mathbb{R}^n$, where $\varsigma_{t_0}=\psi^{-1}[\sigma_{t_0}]$. Note that by construction  $(\mathscr{U}_{[\sigma_{t_0},\,\varepsilon]})\hskip-0.5mm\interior{\phantom{a}}$ is a proper subset of $\widetilde{\mathcal{U}}$, also that $\psi$ is one-to-one on $\mathcal{U}=(\psi[\mathscr{U}_{[\sigma_{t_0},\,\varepsilon]}])\hskip-0.5mm\interior{\phantom{a}}$. Denote by $\phi$ the restriction of the inverse of $\psi$ to $\mathcal{U}=(\psi[\mathscr{U}_{[\sigma_{t_0},\,\varepsilon]}])\hskip-0.5mm\interior{\phantom{a}}$, i.e.,  $\phi=\psi^{-1}|_{(\psi[\mathscr{U}_{[\sigma_{t_0},\,\varepsilon]}])}\hskip-0.5mm\interior{\phantom{{}_a}}$. The  map  $\phi: \mathcal{U} \rightarrow \widetilde{\mathcal{U}}$ is an  embedding that will be used in constructing the desired intermediate extension $\phi: (\mathcal{U},g_{ab}\vert_{\mathcal{U}}) \rightarrow (\widetilde{\mathcal{U}},\widetilde{g}_{ab})$. 

%\bigskip

%\centerline{\includegraphics[width=11cm]{../TEX/eload/2021/EREP-2021/abrak/ext1.pdf}}

%\bigskip

\medskip

Note that, by the above choices made for $\mathcal{U}$ and $\widetilde{\mathcal{U}}$, the members of $\mathcal{G}$ in $\mathcal{U}$ are represented by straight coordinate lines in $\phi[\mathcal{U}]$, and also that for any member $\overline\gamma$ of $\mathcal{G}$, that starts at $\sigma_{t_0}$, and that is future directed incomplete and future inextendible in $(M,g_{ab})$, the curve $\phi\circ\overline\gamma$ can be continued as a straight line in the region $\widetilde{\mathcal{U}}\setminus \phi[\mathcal{U}]$.\,\footnote{Note that in our ultimate argument, in proving Theorem \ref{theor: glob-ext}, $(M,g_{ab})$ is assumed to be a Cauchy development. Whence, if there exists a non-empty boundary to $M$, it has to be part of a Cauchy horizon $\mathcal{H}$ of $(M,g_{ab})$. In particular, the union of the future endpoints of the images of the members of $\mathcal{G}$ in $\widetilde{\mathcal{U}}$, denote it by $H^+$, is a subset  $H^+=\mathcal{H}^+\vert_{\mathcal{U}}$ of the (non-empty) future Cauchy horizon $\mathcal{H}^+$ of $(M,g_{ab})$. Recall then that, in virtue of Proposition 6.3.1, along with the arguments in Section 6.5, of \cite{he}, $H^+ \subset \partial({\phi[\mathcal{U}]})$ is a closed, embedded, achronal three-dimensional $C^{1-}$ submanifold in $\widetilde{\mathcal{U}}$.}

\subsection{Extending the metric from $\mathcal{U}$  to $\widetilde{\mathcal{U}}$}

We shall need the following proposition in extending the metric from $\mathcal{U}$ to $\widetilde{\mathcal{U}}\subset \mathbb{R}^n$. 
\begin{proposition}\label{prop: uniformly continuous}
	Assume that $\phi[\mathcal{U}]$ is defined as above and that it has the property $\mathscr{P}$. Consider a smooth function $\mathcal{F}$ on $\phi[\mathcal{U}]$ and assume that its  first-order partial derivatives, $\partial_t\mathcal{F}$ and $\partial_{x^\alpha}\mathcal{F}$, where $\alpha=1,2,\dots(n-1)$, 
	are uniformly  bounded on $\phi[\mathcal{U}]$. Then, the unique continuous extension $\widetilde{\mathcal{F}}$ of $\mathcal{F}$ to the closure $\overline{\phi[\mathcal{U}]}$ is Lipschitz function that can be further extended onto $\widetilde{\mathcal{U}}\setminus\overline{\phi[\mathcal{U}]}$ such that $\widetilde{\mathcal{F}}$ is also Lipschitz throughout $\widetilde{\mathcal{U}}$.
\end{proposition}
{\noindent \bf Proof:} 
First, the uniform boundedness of the  first-order partial derivatives, $\partial_t\mathcal{F}$ and $\partial_{x^\alpha}\mathcal{F}$,  $\alpha=1,2,\dots(n-1)$, can be used to show that $\mathcal{F}$ is Lipschitz function on $\phi[\mathcal{U}]$ as it was done in proving Proposition 3.3.1 in \cite{racz-1}. Note that then $\mathcal{F}$ is also uniformly continuous there.

The proof of Proposition 4.2 in \cite{racz-2} can be used to verify then that the unique continuous extension $\widetilde{\mathcal{F}}$ of $\mathcal{F}$ onto the closure $\overline{\phi[\mathcal{U}]}$ is also Lipschitz function, with the same Lipschitz constant. 

Finally, in virtue of Kirszbraun’s theorem \cite{Kirszbraun}  $\widetilde{\mathcal{F}}$ also extends to $\widetilde{\mathcal{U}}\setminus\overline{\phi[\mathcal{U}]}$ as a Lipschitz function with the same Lipschitz constant.
{\hfill$\Box$}

\medskip

Note that as $\widetilde{\mathcal{F}}$ is Lipschitz everywhere on $\widetilde{\mathcal{U}}$ the weak derivatives of $\widetilde{\mathcal{F}}$ exist, and they are bounded, whence they are also locally square integrable there.

\bigskip

In proceeding recall first that the $(n-1)\times (n-1)$ matrix elements $g_{\alpha\beta}$ in \eqref{let} are given by the contractions 
\begin{equation}
	g_{\alpha\beta}= g_{ab} E^a_{(\alpha)} E^b_{(\beta)}\,,
\end{equation}
where $E^a_{(\alpha)}$ stand for the coordinate basis elements $(\partial/\partial x^\alpha )^a$, with $\alpha=1,\dots,n-1$. In virtue of Proposition \ref{prop: uniformly continuous}, the components of  $g_{\alpha\beta}$, that are smooth functions on $\phi[\mathcal{U}]$, extend as Lipschitz functions onto $\overline{\phi[\mathcal{U}]}$ if the time- and spatial-derivatives, 
\begin{align}
	\partial_t g_{\alpha\beta} & = g_{ab} \left[\left(v^e\nabla_e E^a_{(\alpha)}\right)  E^b_{(\beta)} + E^a_{(\alpha)} \left( v^e\nabla_e  E^b_{(\beta)}\right)\right] \label{eq: t-deriv} \\ 
	\partial_{x^\nu} g_{\alpha\beta} & = g_{ab} \left[\left(E^e_{(\nu)}\nabla_e E^a_{(\alpha)}\right)  E^b_{(\beta)} + E^a_{(\alpha)} \left( E^e_{(\nu)}\nabla_e  E^b_{(\beta)}\right)\right]\,, \label{eq: x-deriv}
\end{align}
can be guaranteed to be  uniformly bounded along the members of $\mathcal{G}$.

\medskip

Inspecting the individual terms on the right hand sides in \eqref{eq: t-deriv} and \eqref{eq: x-deriv} it appears that it is completely satisfactory to show that the norms $\|E^a_{(\alpha)}\|$, $\|v^e\nabla_e E^a_{(\alpha)}\|$ and  $\|E^e_{(\nu)}\nabla_e E^a_{(\alpha)}\|$ of the vector fields $E^a_{(\alpha)}$, $v^e\nabla_e E^a_{(\alpha)}$ and $E^e_{(\nu)}\nabla_e E^a_{(\alpha)}$ are uniformly bounded with respect to the synchronized orthonormal basis fields $\{e_{(\mathfrak{a})}^a \}$ defined along the members of $\mathcal{G}$ in $\mathcal{U}$. Here the norm $\|X^a\|$ of a vector field $X^a$, with respect to the synchronized basis field $\{e_{{(\mathfrak{a})}}^a \}$ and the Lorentzian metric $g_{ab}$ on $\mathcal{U}$, is defined as 
\begin{equation}
	\|X^a\|:=\sqrt{\sum_{\mathfrak{b}=1}^4 \left[g_{ab}X^a
		e_{{(\mathfrak{b})}}^b \right]^2}\,. 
\end{equation} 
Note that, by applying a straightforward adaptation of Corollary 3.3.5.\;of \cite{racz-1}, the norm  $\|E^a_{(\alpha)}\|$ can be guaranteed to be bounded on $\mathcal{U}$ provided that the tidal force components of the curvature tensor, $R_{abcd}\,e_{{{(\mathfrak{a})}}}^a v^b
e_{{{(\mathfrak{b})}}}^c v^d$, ---measured with respect to a parallelly propagated synchronized orthonormal frame field along the members of $\mathcal{G}$--- are uniformly bounded along the members of $\mathcal{G}$.
Recall also that in a Gaussian coordinate system the coordinate basis fields $E^a_{(\alpha)}=(\partial/\partial x^\alpha )^a$, by construction, are subject to the the Jacobi
equation \eqref{jac}. Applying then Lemma 3.3.6.\;of \cite{racz-1}, to the individual members of $\mathcal{G}$, we get
\begin{equation}
	\|v^e\nabla_e E^a_{(\alpha)}\|_{\overline{\gamma}(t)} \leq \|v^e\nabla_e
	E^a_{(\alpha)}\|_{\overline{\gamma}(t_0)} +\int_{t_0}^t\| {R_{bcd}}^a v^b
	E^c_{(\alpha)} v^d\|_{\overline{\gamma}(t')}\,dt'\,.
\end{equation}
Combining this with the linearity of the integrand in $E^a_{(\alpha)}$, we get that both of the terms $\|E^a_{(\alpha)}\|$ and $\|v^e\nabla_e E^a_{(\alpha)}\|$ are uniformly bounded along the members of $\mathcal{G}$ whenever the tidal force components of the Riemann tensor ---defined with respect to a synchronized orthonormal basis field--- are also guaranteed to remain uniformly bounded along the members of $\mathcal{G}$. 

The characterization of the norm $\|E^e_{(\nu)}\nabla_e E^a_{(\alpha)}\|$ of the vector field $E^e_{(\nu)}\nabla_e E^a_{(\alpha)}$ requires a bit more care. 

Using the linearity of the curvature terms, along with the Leibniz rule, and the vanishing of the commutator of the coordinate basis fields $E^a_{(\alpha)}$ and $v^a$, we get\,\footnote{
Equation \eqref{eq: gen-jac} is a compact form of the generalized Jacobi equation that was introduced in \cite{racz-1} (see also \cite{racz-2}) to characterize the propagation of various order of covariant derivatives of the coordinate basis fields along members of causal geodesic congruences.}
\begin{align}
	v^e\nabla_e(v^f\nabla_f [E^h_{(\nu)}\nabla_hE^a_{(\alpha)}]) = {} & - E^k_{(\nu)}\nabla_k[{R_{efh}}^a v^e E^f_{(\alpha)} v^h] +  {R_{efh}}^a	E^e_{(\nu)} v^f [v^k\nabla_kE_{(\alpha)}^h]  \nonumber \\
	& + v^k\nabla_k[{R_{efh}}^a E_{(\nu)}^e v^f	E_{(\alpha)}^h]  \,. \label{eq: gen-jac}
\end{align}
This, along with
\begin{equation}
	E^a_{(\alpha)}=\sum_{\mathfrak{i}=1}^{n-1} \,\big(g_{kl} E^k_{(\alpha)}\,e_{{{(\mathfrak{i})}}}^l\big) \,e_{{{(\mathfrak{i})}}}^a\,,
\end{equation}
(which follows from the orthogonality of $E^a_{(\alpha)}$ and $v^a$), yields
\begin{align}
\frac{d^2}{dt^2} \,\big(g_{kl} E^k_{(\alpha)}\,e_{{{(\mathfrak{i})}}}^l\big) {} & = - \sum_{\mathfrak{j}=1}^{n-1} 
\big[{R_{abcd}}\, v^a \,e_{{{(\mathfrak{j})}}}^b v^c \,e_{{{(\mathfrak{i})}}}^d\big]\,\big(g_{kl} E^k_{(\alpha)}\,e_{{{(\mathfrak{j})}}}^l\big)\nonumber \\
& \hskip-.5cm  + \sum_{\mathfrak{j}=1}^{n-1} \mathscr{A}_{\mathfrak{j}} \, \big[{R_{abcd}}\,  v^a \,e_{{{(\mathfrak{j})}}}^b v^c \,e_{{{(\mathfrak{i})}}}^d\big] 
+ \sum_{\mathfrak{j,h}=1}^{n-1} \bigg\{\mathscr{B}_{\mathfrak{j}\mathfrak{h}}\,  \big[{R_{abcd}}   \,e_{{{(\mathfrak{i})}}}^a\,e_{{{(\mathfrak{h})}}}^b\,e_{{{(\mathfrak{j})}}}^c v^d \big] \Big. \nonumber \\
& \hskip-.5cm +
\Big. \mathscr{C}_{\mathfrak{j}\mathfrak{h}}\,  \frac{d}{dt}\big[{R_{abcd}}   \,e_{{{(\mathfrak{i})}}}^a\,e_{{{(\mathfrak{h})}}}^b\,e_{{{(\mathfrak{j})}}}^c v^d \big] 
+\mathscr{D}_{\mathfrak{j}\mathfrak{h}}\, e_{(\mathfrak{h})}^k\nabla_k\big[{R_{abcd}}\, v^a \,e_{{{(\mathfrak{j})}}}^b v^c \,e_{{{(\mathfrak{i})}}}^d\big]
\bigg\}  \,, \label{eq: gen-jac-II}
\end{align}
where the coefficients $\mathscr{A}_{\mathfrak{i}}$, $\mathscr{B}_{\mathfrak{i}\mathfrak{j}}$, $\mathscr{C}_{\mathfrak{i}\mathfrak{j}}$ and $\mathscr{D}_{\mathfrak{i}\mathfrak{j}}$  depend exclusively on the fields $E^a_{(\alpha)}$, $v^f\nabla_f E^a_{(\alpha)}$ and $E^f_{(\alpha)}\nabla_f \,e_{{{(\mathfrak{i})}}}^a$. Note that the uniform boundedness of $\|E^a_{(\alpha)}\|$ and $\|v^f\nabla_f E^a_{(\alpha)}\|$ has already been guaranteed by uniform boundedness of the tidal forces. The uniform boundedness of $\|E^f_{(\alpha)}\nabla_f \,e_{{{(\mathfrak{i})}}}^a\|$, however, requires the uniform boundedness of the line integral of the frame-drag part of the curvature. To see this, note that because of the commutation of the coordinate basis fields $E^a_{(\alpha)}$ and $v^a$, also as the basis fields $e_{{{(\mathfrak{i})}}}^a$, $\mathfrak{i}=1,2,\dots(n-1)$, are parallelly propagated with respect to $v^a$, 
\begin{equation}
	v^e\nabla_e (E^f_{(\alpha)}\nabla_f \,e_{{{(\mathfrak{i})}}}^a) = -R_{efh}{}^a v^eE^f_{(\alpha)} e_{{{(\mathfrak{i})}}}^h
\end{equation}
holds, which implies
\begin{equation}
	\frac{d}{dt}\, \big[\,g_{kl}(E^f_{(\alpha)}\nabla_f \,e_{{{(\mathfrak{i})}}}^k)\,e_{{{(\mathfrak{j})}}}^l\,\big] = \sum_{\mathfrak{h}=1}^{n-1} \,[\,g_{kl}E^k_{(\alpha)}\,e_{{{(\mathfrak{h})}}}^l\,] \,[\,R_{abcd} e_{{{(\mathfrak{i})}}}^a e_{{{(\mathfrak{j})}}}^b e_{{{(\mathfrak{h})}}}^c v^d\,]\,.
\end{equation}
Accordingly,
the norm of the $\|E^e_{(\nu)}\nabla_e E^a_{(\alpha)}\|$ of the vector field $E^e_{(\nu)}\nabla_e E^a_{(\alpha)}$ will be uniformly bounded along the members of $\mathcal{G}$ in $\mathcal{U}$ whenever the the tidal force and frame-drag parts of the curvature\,\footnote{The frame-drag part of the curvature gets involved via the line integral of various terms on the right-hand side in  \eqref{eq: gen-jac-II}. However, in the fourth term, the first order $t$-derivative of the frame-drag part is involved, which, in turn, leads to an algebraic restriction on this part of the curvature.}\label{footnote: fn} and the line integrals of the  first-order transversal covariant derivatives of the tidal forces 
\begin{equation}\label{eq: line-int}
	\int_{t_0}^{t}\left(e_{{{(\mathfrak{c})}}}^k\nabla_k[R_{abcd}\,e_{{{(\mathfrak{a})}}}^a v^b
	e_{{{(\mathfrak{b})}}}^c v^d]\right)\vert_{\overline{\gamma}(t')}\,{\rm d}t'\,,
\end{equation}
are all guaranteed to be uniformly bounded along the members of $\mathcal{G}$ in $\mathcal{U}$, where the indices $\mathfrak{a},\mathfrak{b},\mathfrak{c}$ take the values $1,2,\dots,n-1$. 

\medskip 

The critical point here is that it suffices to restrict merely the line integrals of the  first-order transversal covariant derivatives of the tidal forces which follow from the discussion in \cite{racz-2}, starting below (4.15), including (4.16) and the paragraph below there.\,\footnote{One has to consider here the line integrals of the last four terms on the right-hand side in \eqref{eq: gen-jac-II}. Nevertheless, as discussed in footnote \ref{footnote: fn}, the line integral of the second, third and fourth terms leads to an algebraic restriction on the frame-drag part of the curvature. Indeed, it is the fifth term that yields restrictions on the line integrals of the first-order transversal covariant derivatives of the tidal forces.} Note also that the finiteness of these line integrals does not require the integrands' boundedness. They can be uniformly bounded if the integrands do not blow up faster than $(t_p-t)^{-1+\epsilon}$, for some $\epsilon>0$, where $t_p$ stands for the supremum of the affine parameter value along ${\overline{\gamma}}\in \mathcal{G}$.

\medskip 

Utilizing all the above observations, we get that both the time- and spatial-partial derivatives,
$\partial_{t} g_{\alpha\beta}$ and $\partial_{x^\nu} g_{\alpha\beta}$ of the metric will be uniformly bounded along the members of $\mathcal{G}$ whenever the tidal force and frame-drag parts of the curvature, along with the line integrals of the  first-order transversal covariant derivatives of the tidal forces, are guaranteed to be uniformly bounded along the members of $\mathcal{G}$.

\medskip 

If this happens, combining all the above observations, it follows then that the $g_{\alpha\beta}$ components of the metric $g_{ab}$, in Gaussian coordinates,  are Lipschitz functions throughout $\overline{\phi[\mathcal{U}]}$. Proposition \ref{prop: uniformly continuous} also implies that the metric tensor components ${g}_{\alpha\beta}$, in \eqref{let}, extend from $\overline{\phi[\mathcal{U}]}$ onto the entire of $\widetilde{\mathcal{U}}=\varsigma_{t_0}\times[t_0,t_2+\varepsilon)$ such that the extensions $\widetilde{g}_{\alpha\beta}$ are guaranteed to be Lipschitz functions throughout $\widetilde{\mathcal{U}}$.

\medskip 

Note also that the extended metric $\widetilde{g}_{\alpha\beta}$ is $C^0$ Geroch-Traschen regular metric on $\widetilde{\mathcal{U}}$. Firstly, since the components $\widetilde{g}_{\alpha\beta}$ are guaranteed to be Lipschitz functions in $\widetilde{\mathcal{U}}$ the boundedness (not merely the local boundedness) of $\widetilde{g}_{\alpha\beta}$ and that of $\widetilde{g}^{\alpha\beta}$ immediately follows. For the same reason, as the components $\widetilde{g}_{\alpha\beta}$ of the metric are Lipschitz functions on $\widetilde{\mathcal{U}}$ each of the weak derivatives of $\widetilde{g}_{\alpha\beta}$ are well-defined and bounded, thereby, they are also square integrable. This completes the proof of the following: 
\begin{proposition}
	Consider the embedding $\phi: \mathcal{U} \rightarrow \widetilde{\mathcal{U}}$ defined as above. Assume that the tidal force and frame-drag parts of the curvature tensor ---measured with respect to a parallelly propagated synchronized orthonormal frame field along the members of $\mathcal{G}$--- , along with the line integrals of the first-order transversal covariant derivatives of the tidal forces, are uniformly bounded along the members of $\mathcal{G}$. Then there exist an  extension $\phi: (\mathcal{U},g_{ab}\vert_{\mathcal{U}}) \rightarrow (\widetilde{\mathcal{U}},\widetilde{g}_{ab})$  such that $\widetilde{g}_{ab}$ belongs to the $C^0$ Geroch-Traschen regularity class. 
\end{proposition}
Applying the auxiliary extension $\phi: (\mathcal{U},g_{ab}\vert_{\mathcal{U}}) \rightarrow (\widetilde{\mathcal{U}},\widetilde{g}_{ab})$, the desired global extension can then be given as follows: Chose $\mathcal{U}'$ to be a compact subset of $\mathcal{U}$ in $M$. The differentiable structure of $\mathcal{U}$ induces a manifold with boundary on $\mathcal{U}'$. Let, furthermore, $\widetilde{\mathcal{O}}\subset \widetilde{\mathcal{U}}$ be comprised by the union of $\phi[\mathcal{U}']$ and a sufficiently small neighborhood of the endpoint of $\phi\circ\gamma$ in $\widetilde{\mathcal{U}}\subset \mathbb{R}^n$. Define then $\widehat{M}$ to be the factor space 
\begin{equation}
	\widehat{M}=(M\cup\widetilde{\mathcal{O}})/\phi\,,
\end{equation}
i.e., $\widehat{M}$ is yielded by identifying $x\in\mathcal{U}'$ and $y\in \widetilde{\mathcal{O}}$ if $\phi(x)=y$. Then $\widehat{M}$ has the structure of a (Hausdorff) manifold without boundary while the spacetime $(\widehat{M},\widehat{g}_{ab})$ is a $C^0$ extension of $(M,g_{ab})$, where the metric $\widehat{g}_{ab}$ is determined  on $\widehat{M}$ by $g_{ab}$ and $\widetilde{g}_{ab}$, and, thereby, it belongs to the $C^0$  Geroch-Traschen regularity class. 

\medskip
 
Combining all the partial results outlined above verifies then that in the case of a smooth geodesically incomplete generic globally hyperbolic spacetime can be extended within the class of $C^0$ Geroch-Traschen regular metrics if the tidal force and frame-drag parts of the curvature tensor, along with the line integrals of the  first-order transversal covariant derivatives of the tidal forces ---defined for a synchronized basis field---  are guaranteed to be uniformly bounded along the members of $\mathcal{G}$, which completes the proof of Theorem \ref{theor: glob-ext}. 
{\hfill$\Box$}

\section{Final remarks}\label{sec: final}

The two main results of this paper are intimately related. First, ``generic'' (locally algebraically non-special) smooth globally hyperbolic timelike geodesically incomplete spacetime $(M,g_{ab})$ were considered. We showed then that such a spacetime could (globally) be extended within the $C^0$ Geroch-Traschen regularity class if the tidal force and frame-drag parts of the curvature tensor, along with the line integrals of the first-order transversal covariant derivatives of the tidal forces, are uniformly bounded along the world-lines of an $(n-1)$-parameter family of synchronized observers.

\medskip

The second result essentially aims to over-bridging the gap between the intuitive picture of spacetime singularities and the predictions of the singularity theorems. This is made by considering the ``generic'' smooth globally hyperbolic timelike geodesically incomplete spacetime that is a maximal Cauchy development, and assuming that the $C^0$ form of the strong cosmic censor hypothesis holds. Combining this with the first result's conclusion immediately drives to a contradiction, which allows to conclude that there must exist worldlines of observers in arbitrarily small neighborhoods of the one represented by $\gamma$, such that either some of the tidal force or frame-drag components of the curvature, or some of the  first-order transversal covariant derivatives of the tidal force components blow up, at least at the order $(t_p-t)^{-\epsilon}$, for some $\epsilon>0$, or $(t_p-t)^{-1}$, respectively, along the corresponding worldlines of observers.

\medskip

It is worth emphasizing that tidal force and frame-drag parts of the curvature tensor are among the physically most adequate quantities in characterizing spacetime singularities. To see this, note first that the applied $(n-1)$-parameter family timelike geodesics do represent the history of a free-falling small body. Whenever tidal force part of the curvature become unbounded along a member of such a congruence, the small body may undergo an infinite pull-apart in one direction, whereas compression in another. Similarly, one would expect that the blowing up of the integrand in \eqref{eq: line-int} involving  first-order transversal covariant derivatives of the tidal forces, implies that significant shears will be exerted on the internal structure of the aforementioned small body. 
The frame-drag (or magnetic) part of the curvature tensor is also of physical interest as it is directly related to the frame-dragging angular velocity at some event $\bar p\in \overline{\gamma}$, with respect to the inertial directions at a nearby spatially separated event $p\in {\gamma}$. Consider a small body composed of densely arranged tiny gyroscopes. Then, the frame-drag part of the curvature measure the relative precession of the nearby gyroscopes \cite{Nicolson}. Accordingly, the frame-drag part is related to the relative twisting exerted on nearby parts of such a small body, which must undergo infinite wringing if the frame-drag part blows up.

\medskip

Note that mainly for simplicity in this paper, considerations were restricted to the case of timelike geodesics. Nevertheless, the results summarized above generalize straightforwardly (for details, see Refs. \cite{racz-1,racz-2}) to the case when geodesically incomplete lightlike geodesics are involved. 

\medskip

It is worth mentioning that in addition to the geodesic congruence-based approach applied in this paper, there are attempts to give alternative characterizations of spacetime singularities. For instance, there is an approach aiming to characterize spacetime singularities by studying obstructions to the evolution of test fields. The primary motivation behind this approach was that determination of geodesics may become vague in the case of metrics of low regularity \cite{Clark-1998, sanchez, wilson, vickers-wilson}. Note, however, that even the applied low regularity case the method proposed by Clark in \cite{Clark-1998} (see also \cite{wilson,sanchez}) required the construction of congruences of timelike geodesics, whose tangent vectors admit bounded weak derivatives. This vector field was applied to define a suitable energy inequality from which the uniqueness and existence of test fields could be deduced.

\medskip

The last remark immediately raises the following question: Would it be possible to replace the smoothness of the primary metric in Theorems \ref{theor: glob-ext} and \ref{theor: main} by allowing metrics belonging to the class of $C^{0}$ Geroch-Traschen regular metrics? In answering this question, recall, first, that if a Geroch-Traschen regular metric is continuous,\,\footnote{It is worth pointing out that even if a Geroch-Traschen regular metric would not continuous, it could still be approximated by sequences of smooth metrics provided that a specific stability condition, c.f., Section 4 in \cite{Steinbauer-Vickers}, holds on it.}  then it can be approximated by sequences of smooth metrics $\{{}^{(i)}{}g_{ab}\}$ such that the associated curvature tensors $\{{}^{(i)}R_{abc}{}^d\}$ converge in $L^2$ to the curvature distribution assigned to the $C^0$ Geroch-Traschen regular metric $g_{ab}$. To recover the main conclusion in Theorems \ref{theor: glob-ext} and \ref{theor: main}, one has to find a way to represent geodesics with respect to $C^{0}$  Geroch-Traschen regular metrics as an accumulation of sequences of smooth ${}^{(i)\hskip-0.05cm}g_{ab}$-geodesics, as well as, it has to be shown that weak solutions to the Jacobi equation make sense (almost everywhere) along congruences of timelike geodesics.\,\footnote{Note that approximating geodesics of merely continuous or Lipschitz metrics is a subtler matter, since there is no standard way to solve the geodesic equation in these cases. To overcome these difficulties interesting attempts were made by using Fillipov-solutions in case of $C^{0,1}$ regular metrics in \cite{Lange-Lytchak-Samann}\cite{Steinbauer}. Strong results on approximating geodesics of globally hyperbolic $C^1$-metrics were also proved in \cite{Graf}.}
Notably, many of the needed techniques had already been developed and applied in working out the above-mentioned test fields based characterization of spacetime singularities (for details, see \cite{c3}, in particular, the appendix of \cite{sanchez}). 

\medskip

It is also an appealing issue whether the estimates we applied in proving Theorem \ref{theor: glob-ext} were optimal. For instance, it is important to know whether the construction of the intermediate extension could be carried out requiring only the boundedness of the integral

\begin{equation}
	I_{\overline{\gamma}} (t)  = \int_{t_0}^t\| {R_{bcd}}^a v^b E^c_{(\alpha)} v^d \|_{\overline{\gamma} (t)}\,dt\,.
\end{equation}

One of the main obstacles in doing so is that when metrics belonging to the $C^{0}$ Geroch-Traschen regularity class are used, the term ${R_{bcd}}^a v^b E^c_{(\alpha)} v^d$  make sense only as distribution. It is conceivable that this can be done, but it remains to be seen.

%%%%%%%%%%%%%%%%%%%% ACKNOWLEDGMENTS %%%%%%%%%%%%%%%%%%%%%%%%%%%%%%%%%%%
\section*{Acknowledgments}

This project was partly supported by the NKFIH grants K-115434 and K-142423. The author is also indebted to the unknown referees for helpful comments on the previous version of this paper.

%%%%%%%%%%%%%%%%%%%% Data Availability Statement %%%%%%%%%%%%%%%%%%%%%%%%%%%%%%%%%%%
\section*{Data Availability}

Data sharing not applicable to this article as no datasets were generated or analyzed during the current study.

%%%%%%%%%%%%%%%%%%%% Data Availability Statement %%%%%%%%%%%%%%%%%%%%%%%%%%%%%%%%%%%
\section*{Conflict of Interest Statement}

The author declares no conflicts of interest.

%%%%%%%%%%%%%%%%%%%% REFERENCES %%%%%%%%%%%%%%%%%%%%%%%%%%%%%%%%%%%%%%%%
%\section*{References}

\end{document}